\documentclass[12pt,letterpaper]{article}
\usepackage{amsmath,amssymb,amsfonts}
\usepackage{graphicx}
\usepackage{blkarray} 
\usepackage{enumerate}
\usepackage{natbib}
\usepackage{url}

\DeclareMathOperator*{\argmin}{\mbox{argmin}}

\newcommand{\eff}{\mbox{eff}}
\newcommand{\cv}{\mbox{cv}}
\newcommand{\NA}{\mbox{NA}}

\newcommand{\s}{\ell}

\newcommand{\q}{q}

\newcommand{\bw}{\boldsymbol{w}}
\newcommand{\bx}{\boldsymbol{x}}

\newcommand{\bW}{\boldsymbol{W}}
\newcommand{\bX}{\boldsymbol{X}}

\newcommand{\bmu}{\boldsymbol{\mu}}
\newcommand{\btheta}{\boldsymbol{\theta}}
\newcommand{\bSigma}{\boldsymbol{\Sigma}}

\newcommand{\hmu}{\widehat{\mu}}
\newcommand{\hSigma}{\widehat{\Sigma}}

\newcommand{\tw}{\widetilde{w}}
\newcommand{\tW}{\widetilde{W}}

\newcommand{\tmu}{\widetilde{\mu}}
\newcommand{\tSigma}{\widetilde{\Sigma}}

\newcommand{\ttw}{\widetilde{\widetilde{w}}}
\newcommand{\ttW}{\widetilde{\widetilde{W}}}
\newcommand{\ttSigma}{\widetilde{\widetilde{\Sigma}}}

\newcommand{\bhmu}{\boldsymbol{\widehat{\mu}}}
\newcommand{\bhSigma}{\boldsymbol{\widehat{\Sigma}}}

\newcommand{\btmu}{\boldsymbol{\widetilde{\mu}}}
\newcommand{\btSigma}{\boldsymbol{
   \widetilde{\Sigma}}}

\newcommand{\bttSigma}{\boldsymbol{\widetilde{
   \widetilde{\Sigma}}}}

\usepackage[usenames]{color}
\definecolor{blue}{RGB}{0,0,255}
\definecolor{red}{RGB}{255,0,0}

\begin{document}

\def\spacingset#1{\renewcommand{\baselinestretch}
{#1}\small\normalsize} \spacingset{1}


\title{\bf Analyzing cellwise weighted data}
\author{Peter J. Rousseeuw\\
  {\normalsize Section of Statistics and Data 
	Science, University of Leuven, Belgium}\\ \\}
\date{January 3, 2023}
\maketitle

\begin{abstract}
Often the rows (cases, objects) of a dataset
have weights. For instance, the weight of a 
case may reflect the number of times it has 
been observed, or its reliability.
For analyzing such data many rowwise weighted 
techniques are available, the most well known
being the weighted average.
But there are also situations where the
individual {\it cells} (entries) of the data 
matrix have weights assigned to them. 
An approach to analyze such data is proposed.
A cellwise weighted likelihood function is
defined, that corresponds to a transformation
of the dataset which is called unpacking.
Using this weighted likelihood one can carry
out multivariate statistical methods such as
maximum likelihood estimation and likelihood
ratio tests.
Particular attention is paid to the estimation 
of covariance matrices, because these are
the building blocks of much of multivariate
statistics.
An \texttt{R} implementation of the cellwise
maximum likelihood estimator is provided,
which employs a version of the EM algorithm. 
Also a faster approximate method is
proposed, which is asymptotically equivalent
to it. 
\end{abstract}

\vspace{0.5cm} 


\spacingset{1.1} 

\section{Motivation}
\label{sec:intro}

Often the rows (cases, objects) of a dataset
have weights. For analyzing such data 
many rowwise weighted 
techniques are available. For instance,
the concept of a weighted average is widely
known, and has been used extensively in 
areas such as survey sampling.
When the observations are $d$-variate points 
$\bx_1,\ldots,\bx_n$ with nonnegative 
weights $w_1,\ldots,w_n$\,, their weighted
average is simply
\begin{equation} \label{eq:wMean}
 \overline{\bx}_w := 
 \frac{w_1\bx_1 + \ldots + w_n\bx_n}
 {w_1 + \ldots + w_n} \;\;.
\end{equation}
The weight $w_i$ can arise in different ways.
It can be the number of times that $\bx_i$
has been observed (`frequency weight').
But a weight does not have to be an integer: 
the weight $w_i$ can also reflect the 
reliability or precision of the observation 
$\bx_i$\,.
Expression~\eqref{eq:wMean} is also used 
outside of statistics, for instance in physics 
this is the center of gravity of a system 
with masses $w_1,\ldots,w_n$\,.
Note that the effect of the weights 
in~\eqref{eq:wMean} is relative, in the sense
that multiplying all weights by the same 
constant yields the same result.

Similar expressions are those of the weighted 
covariance matrix
\begin{equation} \label{eq:wCov}
 \frac{\sum_{i=1}^n w_i
   (\bx_i - \overline{\bx}_w)
	 (\bx_i - \overline{\bx}_w)^\top}
	 {\sum_{i=1}^n w_i}
\end{equation}
and of weighted least squares regression,
given by
\begin{equation} \label{eq:wLS}
 \argmin_{\btheta}\;
   \sum_{i=1}^n w_i r_i^2
\end{equation}
in which the $r_i$ are the residuals
$y_i - (\theta_0 + \theta_1 x_{i1} +
\ldots + \theta_p x_{ip})$ with the
usual notation. 

All of these can be seen as examples of
maximizing a weighted likelihood.
Let us denote the likelihood of an observation
$\bx$ by $f(\btheta|\bx)$, where the parameter 
$\btheta$ can be a number, vector, matrix etc.
It is often convenient to work with the
loglikelihood 
\begin{equation*}
  L(\btheta|\bx) := \ln f(\btheta|\bx)\;.
\end{equation*}
When the data are independent and identically
distributed (i.i.d.), the loglikelihood of 
the sample $\bX = \{\bx_1,\ldots,\bx_n\}$ is
\begin{equation*}
  L(\btheta|\bX) = \sum_{i=1}^n
	 L(\btheta|\bx_i)\;.
\end{equation*}
The weighted loglikelihood is given by
\begin{equation} \label{eq:cwL}
  L(\btheta|\bX,\bw) = \sum_{i=1}^n
	 w_i L(\btheta|\bx_i)
\end{equation}
where the weights are combined into
the vector $\bw = (w_1,\ldots,w_n)$, and the
corresponding weighted likelihood is
\begin{equation} \label{eq:cwf}
  f(\btheta|\bX,\bw) = \prod_{i=1}^n
	   f(\btheta|\bx_i)^{w_i}\;.
\end{equation}

There is a fairly large literature on the
use of weighted likelihood. 
\cite{Hu1994} and \cite{HuZidek2002} 
consider some data points less relevant than 
others, and wish to diminish their role in
order to trade bias for precision.
The paper by \cite{ohagan2019} focuses in
particular on gaussian mixture models, see 
the function \texttt{me.weighted} in the 
R package \texttt{mclust} \citep{mclust}.
\cite{magis2015} uses weighted likelihood
for item response models.
On the other hand, some authors have used
weighted likelihood to reduce the effect
of potential outliers in estimation, see 
e.g. \cite{fieldSmith1994} for location
and scale, \cite{agostinelliMarkatou1998}
for linear regression, \cite{crouxHR2013}
for ordinal regression, and
\cite{majumder2021} for additional
theoretical properties.
\cite{agostinelliMarkatou2001} focused on
hypothesis tests in this context.

\section{Cellwise weighted likelihood}
\label{sec:weightedL}

The weights we have considered so far were
all rowwise weights, that is, they were
assigned to entire rows of the dataset.
However, it is also possible that the
individual cells (entries) of the data matrix
have weights assigned to them.
For instance, the weight of a cell could be
indicative of the level of confidence in 
that particular measurement, or may be
related to its reliability or measurement
accuracy. It may also be derived from a 
fuzziness measure or a probability.

How can such cellwise weighted data be 
analyzed, that is, how can we estimate 
parameters, carry out tests or other 
inference on them, and make predictions?
As in the previous section we will address
this issue by likelihood, first for a
single row $\bx$.
We assume that each cell of $\bx$ has a 
weight, combined in the weight vector
$$\bw = (w_1,\ldots,w_d)$$
where $w_j \geqslant 0$ for $j=1,\ldots,d$.
A weight $w_j = 0$ is taken to mean that 
the corresponding cell $x_j$ is missing.
Extending our earlier notation, we will
denote the usual observed likelihood 
\citep{LR2020} of a row $\bx$ with some 
missing entries by $f(\btheta|\bx)$ as well.

The question is now whether we can define a 
sensible likelihood in this setting.
First we note that $\bw$ may contain ties,
that is, $w_j = w_{j'}$ for $j \neq j'$.
Let us sort the {\it unique} nonzero weights
as
$$w^{(1)} > w^{(2)} > \ldots > w^{(\q)} > 0$$
with the number of levels $\q \leqslant d$, 
and corresponding sets of indices
$$I^{(\s)} = \{j;\;w_j = w^{(\s)}\} \hspace{1cm}
 \mbox{for} \hspace{1cm} \s=1,\ldots,\q\;.$$
We then consider the cumulative index sets
\begin{align*}
 J^{(1)} &= I^{(1)}\\
 J^{(2)} &= I^{(1)} \cup I^{(2)}\\
 &\hdots\\
 J^{(\q)} &= I^{(1)} \cup I^{(2)} \cup
             \ldots \cup I^{(\q)}
\end{align*}
so $J^{(1)} \subset J^{(2)} \subset
\ldots \subset J^{(\q)}$.
For each $\s=1,\ldots,\q$ we denote by 
$\bx^{(\s)}$ a new row with components
\begin{equation} \label{eq:updateW}
  x_j^{(\s)} = \begin{cases}
  x_j & \mbox{ for }\;\; j \;\; 
	      \mbox{ in }\;\; J^{(\s)}\\
  \mbox{NA} & \mbox{ otherwise } \\
\end{cases}
\end{equation}
for $j=1,\ldots,d$.
We now define the weighted loglikelihood
as the linear combination
\begin{equation} \label{weightedLx}
  L(\btheta|\bx,\bw) := 
	\sum_{\s=1}^{\q}\,
	(w^{(\s)} - w^{(\s+1)})
	L(\btheta|\bx^{(\s)})
\end{equation} 
with the convention $w^{(\q+1)}=0$.
For the likelihood itself this becomes
\begin{equation} \label{weightedfx}
  f(\btheta|\bx,\bw) = 
	\prod_{\s=1}^{\q}\,
	f(\btheta|\bx^{(\s)})^
	{\displaystyle (w^{(\s)}
	- w^{(\s+1)})}\;\;.
\end{equation} 

These formulas look unfamiliar at first, but
when one thinks about it they make perfect
sense. If all $w_j = 1$ one recovers the
usual likelihood, and if all $w_j$ are 0 or
1 it becomes the observed likelihood.
When the cell weights are the number of
repeated measurements, the sets $J^{(\s)}$
are intuitive. 
But as the main benefit we see the ability
to work with the accuracy or trustworthiness
of individual measurements on a continuous
scale.

An i.i.d. sample with $n$ datapoints 
corresponds to an $n \times d$ matrix $\bX$, 
and the weights form an $n \times d$ 
matrix $\bW$.
The overall likelihood of the sample then
becomes
\begin{equation}\label{eq:weightedfX}
  f(\btheta|\bX,\bW) = \prod_{i=1}^n\,
	\prod_{\s=1}^{\q_i}\, 
	f(\btheta|\bx_i^{(\s)})^
	{\displaystyle (w_i^{(\s)}
	- w_i^{(\s+1)})}
\end{equation}
with loglikelihood
\begin{equation} \label{eq:weightedLX}
  L(\btheta|\bX,\bW) = \sum_{i=1}^n\,
	\sum_{\s=1}^{\q_i}\,
	(w_i^{(\s)} - w_i^{(\s+1)})\,
	L(\btheta|\bx_i^{(\s)})\;.
\end{equation}
These formulas have a practical interpretation.
They are equivalent to computing the overall
observed likelihood of an artificial dataset
$\bX^{(\bW)}$ with {\it row} weights, as
in~\eqref{eq:cwL} and~\eqref{eq:cwf}.
The matrix $\bX^{(\bW)}$ is obtained by 
`unpacking' the matrix $\bX$ according to 
the weights in $\bW$.
This is done by replacing each row $\bx_i$
of $\bX$ by $\q_i$ rows $\bx_i^{(\s)}$ that
may contain NA's and have row weights
$v_i^{(\s)}:=(w_i^{(\s)}-w_i^{(\s+1)})>0$.
Rows with a row weight of 0 are left out.
This new matrix $\bX^{(\bW)}$ still has $d$
columns but might have up to $nd$ rows.
When all $w_{ij}=1$ we obtain 
$\bX^{(\bW)} = \bX$, and when all $w_{ij}$
are zero or one we recover the incomplete 
dataset in which the cells $x_{ij}$ with 
$w_{ij}=0$ are set to missing.
Note that the unpacking transform can 
also be used outside of the likelihood
context.

As an illustration, below are
the first 3 rows of a data set $\bX$ with 
four variables, together with the weights
of their cells in the matrix $\bW$:
\begin{equation*}
\bX = 
    \begin{blockarray}{ccccc}
        & 1 & 2 & 3 & 4\\
      \begin{block}{c[cccc]}
         A & 2.8 & 5.3 & 4.9 & 7.4\\
         B & 2.3 & 5.7 & 4.3 & 7.2\\
         C & 2.5 & 5.1 & 4.4 & 7.6\\
	\ldots & \ldots & \ldots & \ldots & \ldots \\
      \end{block}
    \end{blockarray}
\hspace{15mm}
\bW = 		
    \begin{blockarray}{ccccc}
        & 1 & 2 & 3 & 4 \\
      \begin{block}{c[cccc]}
        A & 0.8 & 1.0 & 0.3 & 0.4\\
        B & 0.3 & 0.5 & 0.9 & 0.5\\
        C & 1.0 & 0.6 & 0.0 & 0.7\\
	\ldots & \ldots & \ldots & \ldots & \ldots \\
      \end{block}
    \end{blockarray}\;.	
\end{equation*}

Case A has 4 different nonzero weights, so 
it unpacks into 4 rows of the matrix 
$\bX^{(\bW)}$ below, all labeled as A. 
The first of these rows has the real value 
5.3 in its second position, corresponding 
to the only cell in $\bX$ with weight 
$w_{1j} \geqslant 1.0$, and NA's elsewhere.
Since the next cell will come in at weight
$0.8$, this first row of $\bX^{(\bW)}$ gets
the row weight\linebreak 
 $v_1^{(1)} = 1.0 - 0.8 = 0.2$ 
that we see in the column vector on the right
hand side.
The second row of $\bX^{(\bW)}$ has real
values in cells 1 and 2, which are the cells
of $\bX$ with $w_{1j} \geqslant 0.4$ so the 
weight of this row is 
$v_1^{(2)} = 0.8 - 0.4 = 0.4$ on the right.
The third row has three real values, and 
the fourth row has real values in all of its 
cells.

Next we unpack row B of $\bX$, which is 
analogous except that cells 2 and 4 have the
same cell weight $w_{22} = 0.5 = w_{24}$ so 
there are only three different weights, hence 
row B only yields three rows in $\bX^{(\bW)}$.
Indeed, in row 6 of $\bX^{(\bW)}$ the
entries $5.7$ and $7.2$ join at the same time.
Finally, row C of $\bX$ does have four different
cell weights, but the lowest of them is zero.
The latter would yield a row of $\bX^{(\bW)}$
consisting exclusively of NA's, but such
uninformative rows are not kept, so C also
unpacks into only three rows of $\bX^{(\bW)}$.

\begin{equation*}
\bX^{(\bW)} = 
    \begin{blockarray}{ccccc}
        & 1 & 2 & 3 & 4\\
      \begin{block}{c[cccc]}
A & \NA & 5.3 & \NA & \NA\\ 
A & 2.8 & 5.3 & \NA & \NA\\
A & 2.8 & 5.3 & \NA & 7.4\\
A & 2.8 & 5.3 & 4.9 & 7.4\\
B & \NA & \NA & 4.3  &\NA\\
B & \NA & 5.7 & 4.3 & 7.2\\
B & 2.3 & 5.7 & 4.3 & 7.2\\
C & 2.5 & \NA & \NA & \NA\\
C & 2.5 & \NA & \NA & 7.6\\
C & 2.5 & 5.1 & \NA & 7.6\\
	\ldots & \ldots & \ldots & \ldots & \ldots \\
      \end{block}
    \end{blockarray}
	\hspace{3mm}
		\begin{blockarray}{c}
         v\\
      \begin{block}{[c]}
         0.2\\
				 0.4\\
				 0.1\\
				 0.3\\
				 0.4\\
				 0.2\\
				 0.3\\
				 0.3\\
				 0.1\\
				 0.6\\
		\ldots \\
      \end{block}
    \end{blockarray}\;.
\end{equation*}

One of the important uses of the likelihood
function is to compute the maximum
likelihood estimator (MLE) of $\btheta$.
In view of the matrix unpacking 
interpretation, this is quite feasible.
All we have to do is to apply maximum
likelihood to the unpacked matrix
$\bX^{(\bW)}$ with its row weights.
We will call this estimator the 
{\it cellwise weighted maximum likelihood
estimator} (cwMLE).

For inference it is useful to know the 
large sample behavior of the estimator.
The exact MLE that minimizes the observed 
likelihood is asymptotically normal under 
regularity conditions that are similar to 
those for complete data, as seen in 
Section 6.1.3 of \cite{LR2020} with 
references to proofs.
Therefore the MLE is also consistent.
Some algorithms for the MLE, such as the
Newton-Raphson algorithm, preserve its
asymptotic normality. 
This is also true for the Fisher scoring
algorithm, see e.g. \cite{jamshidian1999},
\cite{jorgenson2012}, and \cite{Takai2020}.
The formulas for the asymptotic covariance
matrix when using Newton-Raphson or
Fisher scoring are given in Section 9.1
of \cite{LR2020}.

The most popular algorithm for the MLE of
incomplete data is the EM algorithm of
\cite{dempster1977}.
The supplemented EM (SEM) algorithm of
\cite{meng1991} also provides, as a byproduct,
a numerically stable estimate of the
asymptotic covariance matrix of the estimator.

In many situations the observed likelihood 
is hard to compute because it requires 
integration, which prevents a closed form.
When that happens one can approximate the
observed likelihood by Monte Carlo, again
yielding asymptotically normal estimates,
see e.g. \cite{sung2007} and the references 
cited therein. 

Apart from estimation, the cellwise weighted
likelihood can also be used for inference,
for instance by applying a likelihood ratio
test using Wilks' chi-squared theorem.

\section{Covariance from cellwise weighted
   data}
\label{sec:muSigma}

We now apply the technology of the previous
section to the ubiquitous multivariate model
where the data $\bX$ are generated from a
gaussian distribution with unknown parameters
$\bmu$ and $\bSigma$.
We will denote the cellwise weighted MLE
(cwMLE) estimates as $\bhmu$ and $\bhSigma$ 
with entries $\hmu_j$ and $\hSigma_{jk}$\,.
An \texttt{R} implementation is available 
which applies the unpacking transform followed
by a rowwise weighted implementation of
the EM algorithm for location and covariance,
which uses iteration.

For rowwise weights we know we can compute 
the weighted MLE by the explicit 
formulas~\eqref{eq:wMean} and~\eqref{eq:wCov}
of the rowwise weighted mean and the 
rowwise weighted covariance.
For cellwise weights no explicit formulas 
for the cwMLE are possible.
But can we at least come up with simple
explicit expressions that {\it approximate}
the cwMLE?
For estimating $\bmu$ it is natural to
consider a cellwise weighted mean (cwMean) 
$\btmu$ given by
\begin{equation}\label{eq:cwMean}
  \tmu_j := \frac{\sum_{i=1}^n w_{ij}x_{ij}}
	 {\sum_{i=1}^n w_{ij}}
\end{equation}
for $j=1,\ldots,d$ in which each coordinate 
$\tmu_j$ uses a different set of weights.

When estimating $\bSigma$ a natural expression 
for the entry $\tSigma_{jk}$ would be
\begin{equation}\label{eq:cellCov}
  \tSigma_{jk} := \frac{\sum_{i=1}^n
	 w_{ijk}(x_{ij}-\tmu_j)(x_{ik}-\tmu_k)}
	 {\sum_{i=1}^n w_{ijk}}\;.
\end{equation}
(As we are approximating an MLE, there is no
analog of subtracting a degree of freedom 
in the denominator.)
The weight $w_{ijk}$ in~\eqref{eq:cellCov}
depends on both the row number $i$ and the 
variables $j$ and $k$. But how should such 
a weight $w_{ijk}$ be defined?
If we think about the construction of the
cellwise loglikelihood~\eqref{eq:weightedLX}
in section \ref{sec:weightedL} and apply it 
to the estimation of $\Sigma_{jk}$\,,
we note that the components $x_{ij}$ and
$x_{ik}$ are only available together in 
some rows of $\bX^{\boldsymbol{(W)}}$, with 
total weight equal to the lowest of $w_{ij}$
and $w_{ik}$\,. Above that level at least 
one of them becomes NA, so in those terms 
of~\eqref{eq:weightedLX}
row $i$ cannot contribute to the
estimation of $\Sigma_{jk}$\,.
This reasoning suggests using
\begin{equation}\label{eq:cwCov}
  \tw_{ijk} := \min(w_{ij},w_{ik})\;.
\end{equation}
We will call the resulting value 
of~\eqref{eq:cellCov} 
the {\it cellwise weighted covariance} 
(cwCov) and denote it as $\tSigma_{jk}$\,. 
Note that for the diagonal entries
$\tSigma_{jj}$ the weights simply
become $\tw_{ijj} = w_{ij}$\,.  
Formulas~\eqref{eq:cellCov} 
and~\eqref{eq:cwCov} are explicit in 
the original $x_{ij}$ and $w_{ij}$
(no unpacking is required) and
allow for fast computation.  
Due to its entrywise construction the 
combined matrix $\btSigma$ need not be 
positive semidefinite (PSD) in general, 
but we will see that it gives an 
excellent approximation to $\bhSigma$ 
and becomes PSD for increasing sample
size.

A different cellwise weighted covariance
matrix was proposed by \cite{VanAelst2011}.
It also falls in the framework 
of~\eqref{eq:cellCov} but uses the
weight function
\begin{equation}\label{eq:sqrtCov}
  \ttw_{ij} = \sqrt{w_{ij}w_{ik}}\;.
\end{equation}
We will denote the resulting entries by
$\ttSigma_{jk}$ forming the matrix
$\bttSigma$ which we call the 
{\it square root covariance matrix}
(sqrtCov), which also is not
necessarily PSD.
Note that the weights used in the
diagonal entries $\ttSigma_{jj}$ also
become $\ttw_{ijj} = w_{ij}$ so
the diagonals of $\bttSigma$ and
$\btSigma$ coincide, but there is no
obvious relation between their
off-diagonal entries. 

\section{Illustration with imprecise
         data cells}\label{sec:jitter}

We now illustrate the behavior of the
cellwise weighted estimators of $\bmu$
and $\bSigma$ in the previous section.
We start by generating $n$ i.i.d.\!  
data points according to the standard
bivariate normal distribution, so $d=2$.
Next we `jitter' some of the cells in 
the following way.
We randomly draw 20\% of the data cells,
and add independent noise to them that
is normally distributed with mean zero
and standard deviation 3.
An equivalent way to formulate this
jittering scenario is to say that the
data cells $x_{ij}$ all have a univariate
normal distribution with mean zero,
most of them with variance $v_{ij}=1$ 
except for a random fraction of 20\% of 
the cells that has variance 
$v_{ij} = 3^2 +1 = 10$.
The latter cells can be seen as less 
precise than the remaining 80\%.

It is still possible to estimate $\bmu$ 
by the classical mean $\overline{\bx}$,
whose components remain unbiased since
all cells have mean zero.
But $\overline{\bx}$ gives every cell 
the same weight, which does not reflect 
the differences in precision. 
Alternatively we could assign weights
$w_{ij}$ to the cells that are a 
decreasing function of the variances, 
for instance $w_{ij} = 1/v_{ij}^2$\,.

We ran a small simulation, consisting
of 5000 replications for sample sizes
$n$ ranging from 10 to 10000.
Apart from $\overline{\bx}$ and the
classical covariance matrix Cov we
also computed the estimates $\bhmu$
and $\bhSigma$ obtained by cwMLE,
the cwMean vector $\btmu$ and cwCov 
matrix $\btSigma$, and the sqrtCov 
matrix $\bttSigma$. 
For $n \geqslant 20$ both the cwCov
and sqrtCov matrices were positive 
definite in all 5000 replications.
Table~\ref{tab:avejittered} reports
average values of the components of
all these estimates.

\begin{table}[!ht] 		
\centering
\caption{Average of estimates when 
         there are imprecise data cells.}
\label{tab:avejittered}				
\vskip3mm
\footnotesize
\begingroup
\renewcommand{\arraystretch}{0.9} 
\begin{tabular}{rcccccccccc}
\hline
	& \multicolumn{3}{c}{\underline{estimates for the $\mu_j$}}
	& \multicolumn{3}{c}{\underline{for diagonal entries of $\bSigma$}}
	& \multicolumn{4}{c}{\underline{for the off-diagonal entries of $\bSigma$}}\\
 $n$ & $\overline{\bx}$ & cwMLE & cwMean & Cov & 
 cwMLE & cwCov & Cov & cwMLE & cwCov & sqrtCov\\
\hline
   10 & 0.004 & 0.000 & 0.001 & 2.782 & 0.919 &  0.903 & 0.007 & 0.002 & 0.006 & 0.007\\
   20 & 0.005 & 0.004 & 0.004 & 2.802 & 0.973 & 0.970 & -0.006 & 0.001 & 0.002 & 0.001\\
   50 & -0.004 & -0.001 & -0.001 & 2.791 & 0.998 & 0.998 & -0.010 & 0.003 & 0.004 & 0.003\\
  100 & 0.003 & 0.000 & 0.000 & 2.808 & 1.014 & 1.014 & -0.002 & -0.004 & -0.004 & -0.004\\
  200 & 0.000 & 0.000 & 0.000 & 2.803 & 1.017 & 1.017 & -0.001 & 0.002 & 0.002 & 0.002\\
  500 & 0.000 & 0.001 & 0.001 & 2.800 & 1.021 & 1.021 & 0.001 & 0.000 & 0.000 & 0.000\\
 1000 & 0.000 & 0.000 & 0.000 & 2.800 & 1.021 & 1.021 & -0.001 & 0.000 & 0.000 & 0.000\\
 2000 & 0.000 & 0.000 & 0.000 & 2.798 & 1.021 & 1.021 & -0.002 & 0.000 & 0.000 & 0.000\\
 5000 & 0.000 & 0.000 & 0.000 & 2.801 & 1.022 & 1.022 & -0.001 & 0.000 & 0.000 & 0.000\\
10000 & 0.000 & 0.000 & 0.000 & 2.800 & 1.022 & 1.022 & 0.000 & 0.000 & 0.000 & 0.000\\
\hline
\end{tabular}
\endgroup
\end{table}

As expected, we see that all three 
estimators of the $\mu_j = 0$ tend
to zero. Also the off-diagonal entries
of the covariance estimators tend to
zero, which is intuitive due to the
symmetries in the data.
When estimating the diagonal entries
$\Sigma_{jj}$ the situation is quite
different, as the classical Cov goes
to $0.80 + 0.20*10 = 2.8$\,.
The estimators cwMLE and cwCov stay 
much closer to 1 because they 
downweight the imprecise cells.

Table~\ref{tab:varjittered}	shows the
variances of the estimators over the
5000 replications, multiplied by the
sample size $n$.
Here we see that the cellwise weighted
estimators of $\mu_j$ have a much
lower variance than the classical mean,
which attaches the same weight to the
precise and the imprecise cells.
This effect is even more pronounced
for the estimates of the off-diagonal
entries $\Sigma_{jk}$ where the 
variance of the classical covariance 
is inflated more relative to the
cellwise weighted estimators.
Note that we divided the variances of 
the estimates of the diagonal entries 
$\Sigma_{jj}$ by 2, which would be the 
lowest achievable variance if all cells 
were precise. 
In those columns the entries for the 
diagonal of the unweighted Cov are much
higher than those of cwMLE and cwCov
due to the imprecise data cells. 
Overall, the cellwise weighted
estimators performed the best in this
mixed precision setting.

\begin{table}[!ht] 		
\centering
\caption{Variance of estimates when 
         there are imprecise data cells.}
\label{tab:varjittered}				
\vskip3mm
\small
\begingroup
\renewcommand{\arraystretch}{0.9} 
\begin{tabular}{rcccccccccc}
\hline
	& \multicolumn{3}{c}{\underline{estimates for the $\mu_j$}}
	& \multicolumn{3}{c}{\underline{for diagonal entries of $\bSigma$}}
	& \multicolumn{4}{c}{\underline{for the off-diagonal entries of $\bSigma$}}\\
 $n$ & $\overline{\bx}$ & cwMLE & cwMean & Cov & 
 cwMLE & cwCov & Cov & cwMLE & cwCov & sqrtCov\\
\hline
   10 & 2.72 & 1.32 & 1.26 & 24.77 & 1.24 & 1.12 & 8.01 & 1.72 & 1.33 & 1.27\\
   20 & 2.76 & 1.25 & 1.23 & 24.42 & 1.22 & 1.20 & 7.84 & 1.62 & 1.45 & 1.39\\
   50 & 2.75 & 1.24 & 1.23 & 24.45 & 1.22 & 1.22 & 7.71 & 1.56 & 1.51 & 1.46\\
  100 & 2.77 & 1.24 & 1.24 & 24.35 & 1.25 & 1.25 & 8.08 & 1.52 & 1.50 & 1.45\\
  200 & 2.70 & 1.24 & 1.23 & 24.01 & 1.23 & 1.23 & 8.11 & 1.53 & 1.56 & 1.51\\
  500 & 2.74 & 1.23 & 1.23 & 23.62 & 1.23 & 1.23 & 7.78 & 1.52 & 1.56 & 1.50\\
 1000 & 2.76 & 1.25 & 1.25 & 24.13 & 1.24 & 1.24 & 8.08 & 1.52 & 1.55 & 1.49\\
 2000 & 2.79 & 1.26 & 1.26 & 24.03 & 1.24 & 1.24 & 8.03 & 1.52 & 1.55 & 1.50\\
 5000 & 2.83 & 1.22 & 1.22 & 24.21 & 1.27 & 1.27 & 7.97 & 1.50 & 1.54 & 1.47\\
10000 & 2.79 & 1.26 & 1.26 & 24.10 & 1.29 & 1.29 & 7.90 & 1.51 & 1.54 & 1.48\\
\hline
\end{tabular}
\endgroup
\end{table}

\begin{figure}[!ht]
\centering
\vspace{0.2cm}
\includegraphics[width=1.0\textwidth]
  {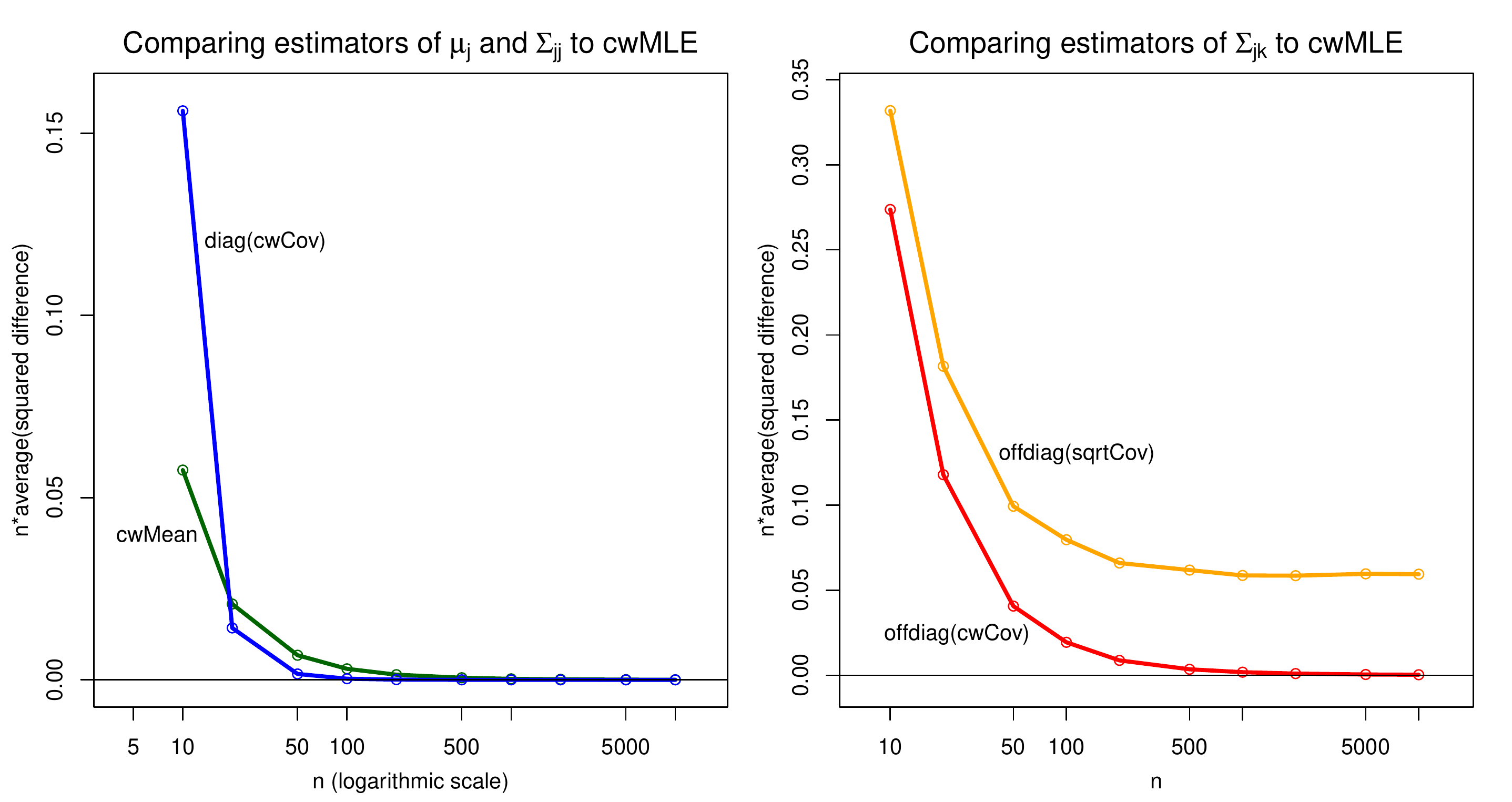}
\vspace{-0.8cm}	
\caption{Data with imprecise cells:
  plot of~\eqref{eq:asyEq} for 
  (left) the estimator cwMean and the
	diagonal entries of cwCov, and (right)
  the off-diagonal entries of cwCov 
	(lower curve) and of sqrtCov 
	(upper curve).}
\label{fig:AsyEqJitter}				
\end{figure}

In Tables~\ref{tab:avejittered}
and~\ref{tab:varjittered} we see that
the entries for $\overline{x}_j$ and
the cwMLE estimator of $\mu_j$ are
close to each other, especially for
large $n$.
In fact, we can see a bit more.
From the simulated estimates we also
computed
\begin{equation}\label{eq:asyEq}
  n\, \frac{1}{Md} \sum_{m=1}^M {
  \sum_{j=1}^d {
	(\hmu_j^{(m)} - \tmu_j^{(m)})^2 }}
\end{equation}
where $\hmu_j^{(m)}$ is the estimate in
replication $m$ for $m=1,\ldots,M$.
The left panel of 
Figure~\ref{fig:AsyEqJitter} shows
this as a function of $n$ in the curve 
labeled cwMean. 
We see that it goes down to zero for
increasing $n$, indicating that 
cwMean is in fact asymptotically 
equivalent to the estimate $\bhmu$ of
cwMLE.
(Note that the Chebyshev inequality
implies that $\sqrt{n}(\tmu_j-\hmu_j)$
goes to zero in probability.)
We see the same effect for the analogous 
quantity comparing the diagonal entries 
of cwCov with $\hSigma_{jj}$\,.
The bottom curve in the right panel of
Figure~\ref{fig:AsyEqJitter} compares
the off-diagonal entries of cwCov
with $\hSigma_{jk}$ and goes to zero too.
All of this suggests that the combination 
of cwMean and cwCov is asymptotically 
equivalent to the cwMLE method, which
is intuitively understandable since the 
construction of the 
weights~\eqref{eq:cwCov} mimics the 
guiding principle of the cellwise 
weighted likelihood.
On the other hand, the upper curve in
the right panel does not go to zero, 
indicating that sqrtCov is not
asymptotically equivalent with cwMLE.

The asymptotic equivalence of the pair 
(cwMean,cwCov) with cwMLE can be 
exploited in several ways. 
Since cwMean and cwCov are easy to 
compute, they could be used as  
replacements or approximations of cwMLE 
when cwCov is PSD. 
And if cwCov is not PSD we can easily
regularize it by carrying out the
spectral decomposition of cwCov and 
replacing its nonpositive eigenvalues 
by a small positive number.

Another possibility is to use cwMean 
and cwCov (regularized when needed) as 
initial estimates in the algorithm of 
cwMLE. This is now an option in the 
\texttt{R} code. 
In experiments this reduced the number 
of iteration steps, while the result 
was identical.
(The simulation yielding 
Figure~\ref{fig:AsyEqJitter} didn't
use this option, so the effect we saw
is not an artifact of the 
computation.) 

\section{Results with precise data cells 
         and random weights}
				 \label{sec:random}
				
We now look at a different situation, where
the data cells are actually precise but we
use weights that are not constant.
We generate i.i.d. data $\bx_i$ following 
a multivariate gaussian random variable 
$X$, and let the weights $w_{ij}$ in the 
matrix $\bW$ be i.i.d. according to a random 
variable W that is independent of $X$. 
The latter condition resembles the missing
completely at random (MCAR) assumption 
for missing data.
In this setting one can verify that the
components $\tmu_j$ of cwMean
are asymptotically normal, and hence 
consistent.
The asymptotic variance of $\tmu_j$ 
equals that of the unweighted mean 
(which is also the unweighted MLE) 
multiplied by the factor
\begin{equation}\label{eq:asyMu}
  V(W) := \frac{E[W^2]}{E[W]^2}
\end{equation}
which is at least 1 since 
$E[W^2] - E[W]^2 = Var[W] \geqslant 0$,
so the asymptotic efficiency  
$\eff = 1/V(W)$ is at most 1.
The same factor $V(W)$ also multiplies
the asymptotic variance of the diagonal
elements 
$\tSigma_{jj} = \ttSigma_{jj}$\,.
The variances of the off-diagonal 
entries $\tSigma_{jk}$
are instead multiplied by $V(\tW)$
where $\tW = \min(W_1,W_2)$ in which
$W_1$ and $W_2$ are independent copies
of $W$. For the sqrtCov matrix the
factor becomes $V(\ttW)$
where $\ttW = \sqrt{W_1 W_2}$\,.

A small simulation was run to illustrate 
these properties.
The data were generated from the bivariate 
standard gaussian distribution, with
5000 replications for each value of $n$.
The weights were randomly generated
according to the uniform random variable 
W on $[0,1]$.
Using the uniform variable $W$ yields
the population factor 
$V(W) = (1/4)/(1/3) = 4/3 \approx 1.33$
for the asymptotic variance of cwMean 
and the diagonal entries of cwCov and
sqrtCov.
For the off-diagonal entries of cwCov
we require the distribution of $\tW$
which has density 
$f(w) = 2(1-\tw)I(0\leqslant \tw
\leqslant 1)$ and $E[\tW]^2 = 1/9$, 
$E[\tW^2] = 1/6$ so 
$V(\tW) = 3/2 = 1.50$\,.
The computation is a bit harder for 
the off-diagonal entries of sqrtCov.
There $\ttW$ has density
$g(\ttw) = 4\ttw\log(1/\ttw)I(0\leqslant 
\ttw \leqslant 1)$ which yields
$E[\ttW]^2 = 16/81$ and 
$E[\ttW^{\,_2}] = 1/4$ 
so $V(\ttW) = 81/64 \approx 1.27$\,.

The entries in Table~\ref{simuniform}
are the mean squared errors of the 
estimates for $\mu_j$ averaged over 
$j=1,2$, and likewise for the 
off-diagonal entries $\Sigma_{jk}$\,.
Since the unweighted MLE estimators
$\overline{\bx}$ and Cov are efficient
for these data with precise cells, we 
do not list them here.
The MSE of the cellwise weighted 
estimates for $\hmu_j$ should trend to 
the value $V(F)$.
For the diagonal entries $\Sigma_{jj}$ 
we divide the MSE by 2 (which is the
asymptotic variance of the unweighted
estimator) so the result should go to
$V(F)$ as well.
For the off-diagonal entries, the MSE
should trend to $V(\tW)$ for cwMLE
and cwCov, and to $V(\ttW)$ for
sqrtCov.

\begin{table}[!ht] 		
\centering
\caption{MSE multiplication factors of cellwise
         weighted estimators when the weights 
				 are random and uniform on $[0,1]$.}
\label{simuniform}				
\vskip3mm
\begingroup
\renewcommand{\arraystretch}{0.9} 
\begin{tabular}{rccccccc}
\hline
	& \multicolumn{2}{c}{\underline{estimates for $\bmu$}}
	& \multicolumn{2}{c}{\underline{for diagonal of $\bSigma$}}
	& \multicolumn{3}{c}{\underline{for off-diagonal of $\bSigma$}}\\
 $n$ & cwMLE & cwMean & cwMLE & cwCov & 
       cwMLE & cwCov & sqrtCov\\
\hline
10 & 1.31 & 1.31 & 1.22 & 1.22 & 1.23 & 1.19 & 1.03\\
20 & 1.33 & 1.34 & 1.26 & 1.25 & 1.36 & 1.30 & 1.11\\
50 & 1.30 & 1.30 & 1.35 & 1.35 & 1.43 & 1.41 & 1.21\\
100 & 1.34 & 1.34 & 1.34 & 1.34 & 1.48 & 1.48 & 1.26\\
200 & 1.35 & 1.35 & 1.32 & 1.32 & 1.49 & 1.50 & 1.26\\
500 & 1.31 & 1.31 & 1.28 & 1.28 & 1.46 & 1.46 & 1.24\\
1000 & 1.35 & 1.35 & 1.31 & 1.31 & 1.48 & 1.48 & 1.26\\
2000 & 1.35 & 1.35 & 1.32 & 1.32 & 1.52 & 1.52 & 1.28\\
5000 & 1.31 & 1.31 & 1.29 & 1.29 & 1.50 & 1.50 & 1.27\\
10000 & 1.34 & 1.34 & 1.37 & 1.37 & 1.47 & 1.47 & 1.25\\
$\infty$ & 1.33 & 1.33 & 1.33 & 1.33 & 1.50 & 1.50 & 1.27\\
\hline
\end{tabular}
\endgroup
\end{table}

In Table~\ref{simuniform} we see
that for $n \geqslant 100$ the empirical
MSE multiplication factors are quite 
close to their population versions
listed in the row $n=\infty$. We also 
note that the MSE values of the cwMLE 
location are close to those of cwMean, 
that those of the diagonal of the cwMLE 
covariance are close to those of cwCov, 
and similarly for the off-diagonal
entries of these covariances. This 
confirms our expectation that the 
asymptotic variances of cwMLE coincide 
with those of cwMean and cwCov.

\begin{figure}[!ht]
\centering
\vspace{0.2cm}
\includegraphics[width=1.0\textwidth]
  {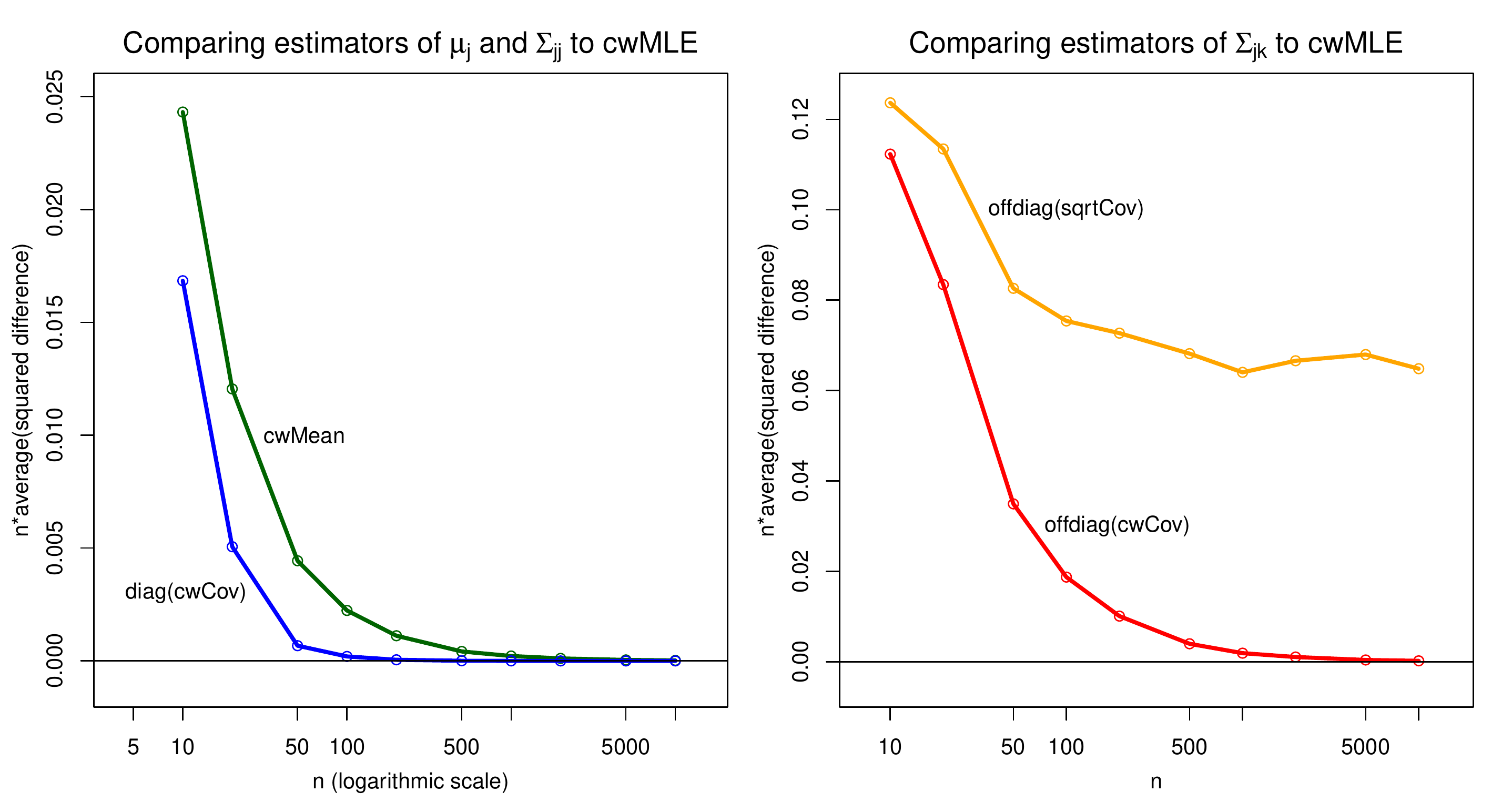}
\vspace{-0.8cm}	
\caption{Precise data cells with random
  weights: plot of~\eqref{eq:asyEq} for 
  (left) the estimator cwMean and the
	diagonal entries of cwCov, and (right)
  the off-diagonal entries of cwCov 
	(lower curve) and of sqrtCov 
	(upper curve).}
\label{fig:AsyEqUniform}				
\end{figure}

The left panel of 
Figure~\ref{fig:AsyEqUniform} 
shows~\eqref{eq:asyEq} as in 
Figure~\ref{fig:AsyEqJitter} and
again indicates that cwMean is 
asymptotically equivalent to the 
estimator $\bhmu$ of cwMLE.
The other curves in
Figure~\ref{fig:AsyEqUniform} 
reflect that cwCov is asymptotically
equivalent to the cwMLE covariance 
estimator whereas sqrtCov is not.

In the last column of 
Table~\ref{simuniform} we see that the
off-diagonal entries of sqrtCov are 
more efficient than those of cwCov.
There are two reasons for this.
First, the simulation is for the
idealized situation where the $\bX$ 
sample is perfectly gaussian with 
constant accuracy, and in that
situation the unweighted covariance
Cov would perform best.
And secondly, sqrtCov is more similar 
to Cov than cwCov is, since its 
weights are closer to constant.
Since weights are only defined up to a
factor, how close they are to constant
can be measured by their coefficient 
of variation 
$\cv[W] = \mbox{Stdev}[W]/E[W]$.
It is easily seen that there is a 
monotone relation between the variance 
factor and the coefficient of 
variation:
\begin{equation}\label{eq:cv}
  V(W) = \mbox{cv}(W)^2 + 1\,.
\end{equation}
In the current setting we have
$\cv(\tW)^2 = 1/2$ and 
$\cv(\ttW)^2 = 17/64 \approx 0.2656$
so the weights of sqrtCov have a lower
cv than those of cwCov, and hence 
yield a lower variance factor. 
For a constant weight $W$ we would get 
$\cv(W) = 0$ yielding a variance 
factor of 1, which is
the lowest possible.

Repeating the simulation for other
dimensions $d$ gave similar results 
(not shown).

Let us now consider the situation where
the dataset $\bX$ contains NA's that
are missing completely at random.
This can be put in our framework by using
a matrix $\bW$ of cell weights that are 0 
or 1, such that the zeroes in $\bW$ are 
placed at the positions of the NA's in $\bX$.
In that situation the unpacked matrix
$\bX^{(\bW)}$ is just $\bX$ and all its
row weights are 1.
Therefore, the cellwise weighted likelihood
coincides with the observed likelihood of
the incomplete dataset $\bX$.
The cwMLE method then reduces to the MLE
of incomplete data, whose computation
requires iteration.
On the other hand, we can still compute
cwMean and cwCov explicitly.
Note that sqrtCov coincides with cwCov
in this setting, because for zero-one
weights $w_{ij}$ and $w_{ik}$ it holds 
that $\min(w_{ij},w_{ik}) =
\sqrt{w_{ij} w_{ik}}$\,.
Also note that in this situation the
entry $\tSigma_{jk}$ of cwCov is just 
the average of the 
$(x_{ij} - \tmu_j)(x_{ik} - \tmu_k)$
over the pairs with both $x_{ij}$ and
$x_{ik}$ non-missing.

We have simulated the MCAR setting by
generating the weights from a Bernoulli 
random variable with success probability
0.9, corresponding to 10\% of missing 
values.
The $\bX$ data were generated as before.
We ran 5,000 replications for each sample
size $n$.
From the properties of the Bernoulli
random variable $W$ we immediately
obtain the variance factor  
$V(W) = E[W^2]/E[W]^2 = 0.9/(0.9)^2
= 1/0.9 \approx 1.11$\,. Since the
distribution of $\tW$ is Bernoulli with
success probability $0.9^2 = 0.81$ we
analogously find 
$V(\tW) = 1/0.81 \approx 1.23$\,. 

\begin{table}[!ht]			
\centering
\caption{MSE multiplication factors of cellwise
         weighted estimators 
         when the weights are zero-one with the
				 zeroes at MCAR missing values.}
\label{simMCAR}
\vskip3mm
\begingroup
\renewcommand{\arraystretch}{0.9} 
\begin{tabular}{rcccccc}	
\hline
	& \multicolumn{2}{c}{\underline{estimates for $\bmu$}}
	& \multicolumn{2}{c}{\underline{for diagonal of $\bSigma$}}
	& \multicolumn{2}{c}{\underline{for off-diagonal of $\bSigma$}}\\
 $n$ & cwMLE & cwMean & cwMLE & cwCov & cwMLE & cwCov\\
\hline	
   10 & 1.14 & 1.10 & 1.14 & 1.10 & 1.36 & 1.15\\
   20 & 1.13 & 1.12 & 1.10 & 1.09 & 1.24 & 1.15\\
   50 & 1.09 & 1.09 & 1.11 & 1.11 & 1.23 & 1.20\\
  100 & 1.09 & 1.09 & 1.13 & 1.13 & 1.25 & 1.24\\
  200 & 1.09 & 1.09 & 1.14 & 1.14 & 1.26 & 1.25\\
  500 & 1.11 & 1.11 & 1.09 & 1.09 & 1.24 & 1.24\\
 1000 & 1.11 & 1.11 & 1.10 & 1.10 & 1.26 & 1.26\\
 2000 & 1.11 & 1.11 & 1.09 & 1.09 & 1.28 & 1.28\\
 5000 & 1.12 & 1.12 & 1.10 & 1.10 & 1.22 & 1.22\\
10000 & 1.13 & 1.13 & 1.11 & 1.11 & 1.24 & 1.24\\
$\infty$ & 1.11 & 1.11 & 1.11 & 1.11 & 1.23 & 1.23\\
\hline
\end{tabular}
\endgroup
\end{table}
\normalsize

\begin{figure}[!ht]
\centering
\includegraphics[width=1.0\textwidth]
  {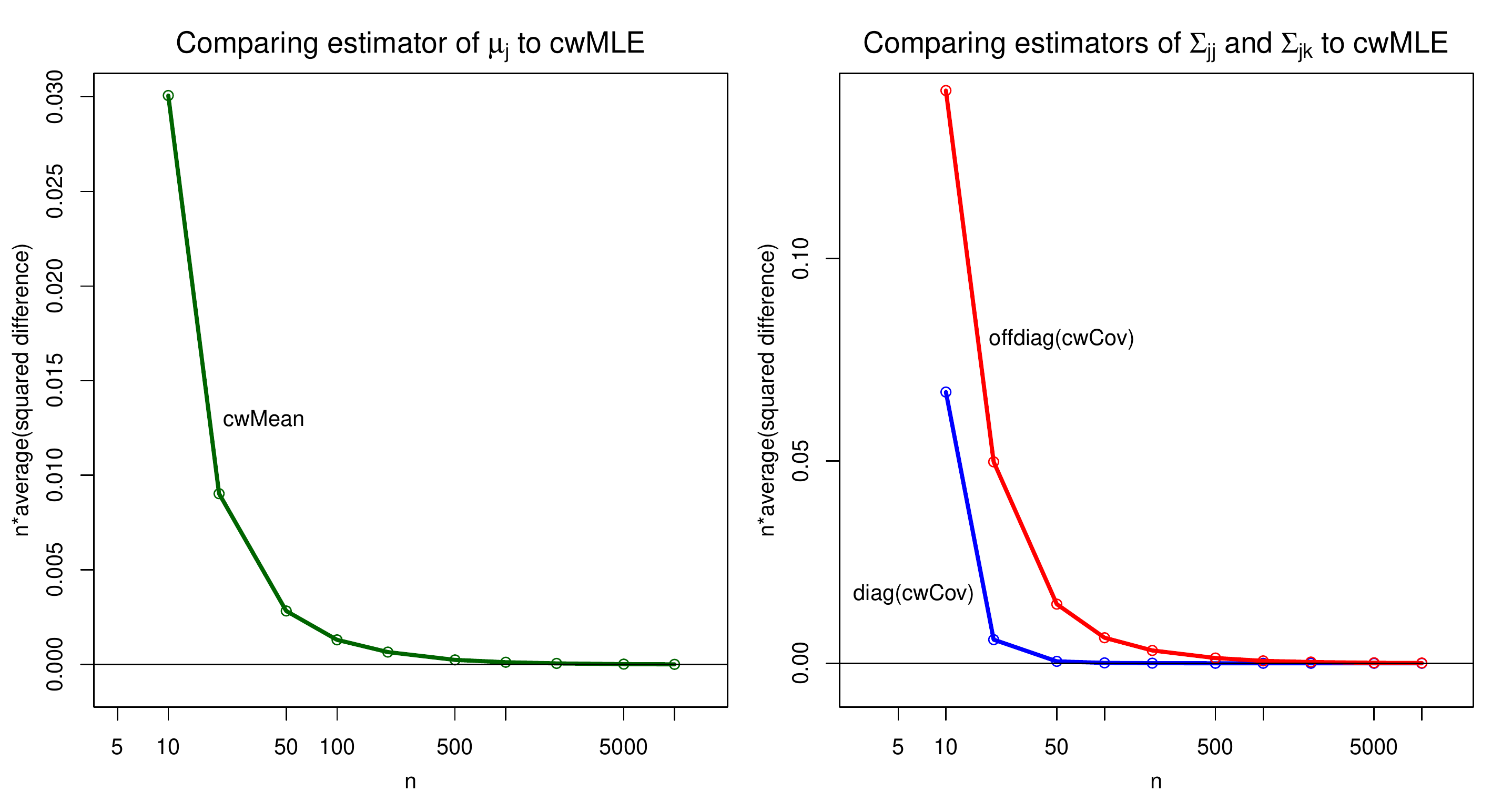}
\vspace{-0.8cm} 
\caption{Data with MCAR missing values: plot 
  of~\eqref{eq:asyEq} for (left) the estimator 
  cwMean, and (right) the diagonal and
  off-diagonal entries of cwCov.}
\label{fig:AsyEqMCAR}
\end{figure}

In Table~\ref{simMCAR} we again see that the
limiting behavior takes hold already at low
sample sizes. Not surprisingly, 
the efficiency of the location 
estimates and the diagonal of the covariance
matrices is $1/V(W) = 90\%$ which is the 
fraction of non-missing cells $x_{ij}$\,.
Analogously, the efficiency of the
off-diagonal of the covariance is
$1/V(\tW) = 81\%$, the percentage of 
non-missing pairs $(x_{ij},x_{ik})$.
Also, Figure~\ref{fig:AsyEqMCAR}
illustrates the asymptotic equivalence of the
combination of cwMean and cwCov with cwMLE.
Since in the MCAR situation cwMLE is just the
usual MLE for incomplete data, and cwCov has
the simple expression above, this asymptotic 
equivalence was presumably known before
to some.

\section{Example} \label{sec:example}

Cellwise weights can be due to varying accuracy or
reliability of entries in the data matrix, which
differs from the concept of random noise that
underlies much of statistics.
Often it is assumed that all cells are equally
accurate, but this may not be true in reality.
A scientific community that cares about the
accuracy of data is that of soft computing, and
in particular fuzzy numbers.
A fuzzy number is a fuzzy set, which is not
localized in a single point but has a membership
function. The more spread out the fuzzy number,
the less accurate the measurement is.

As an example we consider a fuzzy dataset of
\cite{Hesamian2019} about six personality 
traits of 10 subjects. 
The data matrix $\bX$ is in the left panel of
Table~\ref{exampledata}. 
The weights in the right panel are the inverse
of the length of the support of the membership
functions, normalized so the largest weight
equals one.
Due to its small sample size and lack of detail
this dataset is not very interesting in itself, 
but it offers the opportunity to illustrate some
aspects of the methods developed here.

\begin{table}[!ht]			
\centering
\caption{Cellwise weighted data on personality traits.}
\label{exampledata}
\vskip3mm
\begingroup
\renewcommand{\arraystretch}{0.9} 
\begin{tabular}{ccccccccccccc}	
\hline
	\multicolumn{6}{c}{data matrix $\bX$} &
	\multicolumn{1}{c}{\phantom{abc}} &
	\multicolumn{6}{c}{weight matrix $\bW$}\\
  t1 & t2 & t3 & t4 & t5 & t6 & \phantom{abc} & t1 & t2 & t3 & t4 & t5 & t6\\
\hline	  
 7 & 5 & 7 & 5 & 5 & 5 & \phantom{abc} & 0.50 & 0.29 & 0.50 & 0.29 & 0.29 & 0.29\\
 10 & 10 & 10 & 7 & 8.5 & 7 & \phantom{abc} & 1.00 & 1.00 & 1.00 & 0.50 & 0.58 & 0.50\\
 5 & 5 & 10 & 5 & 5 & 5 & \phantom{abc} & 0.29 & 0.29 & 1.00 & 0.29 & 0.29 & 0.29\\
 10 & 10 & 10 & 5 & 5 & 5 & \phantom{abc} & 1.00 & 1.00 & 1.00 & 0.29 & 0.29 & 0.29\\
 7 & 7 & 8.5 & 5 & 5 & 5 & \phantom{abc} & 0.50 & 0.50 & 0.58 & 0.29 & 0.29 & 0.29\\
 10 & 5 & 5 & 8.5 & 8.5 & 5 & \phantom{abc} & 1.00 & 0.29 & 0.29 & 0.58 & 0.58 & 0.29\\
 5 & 7 & 7 & 5 & 5 & 8.5 & \phantom{abc} & 0.29 & 0.50 & 0.50 & 0.29 & 0.29 & 0.58\\
 10 & 10 & 10 & 10 & 10 & 10 & \phantom{abc} & 1.00 & 1.00 & 1.00 & 1.00 & 1.00 & 1.00\\
 8.5 & 7 & 8.5 & 5 & 5 & 5 & \phantom{abc} & 0.58 & 0.50 & 0.58 & 0.29 & 0.29 & 0.29\\
 5 & 10 & 5 & 7 & 5 & 7 & \phantom{abc} & 0.29 & 1.00 & 0.29 & 0.50 & 0.29 & 0.50\\
\hline
\end{tabular}
\endgroup
\end{table}
\normalsize

Estimating the covariance matrix of these 
cellwise weighted data by cwMLE is immediate.
We looked at scatterplots of each pair of 
variables, with the 95\% confidence 
ellipses of cwMLE as well as those of the 
plain unweighted MLE.
In most of these plots the ellipses looked
rather similar, but let us consider a 
pair of variables for which they differ. 

\begin{figure}[!ht]
\centering
\vspace{0.2cm}
\includegraphics[width=0.6\textwidth]
  {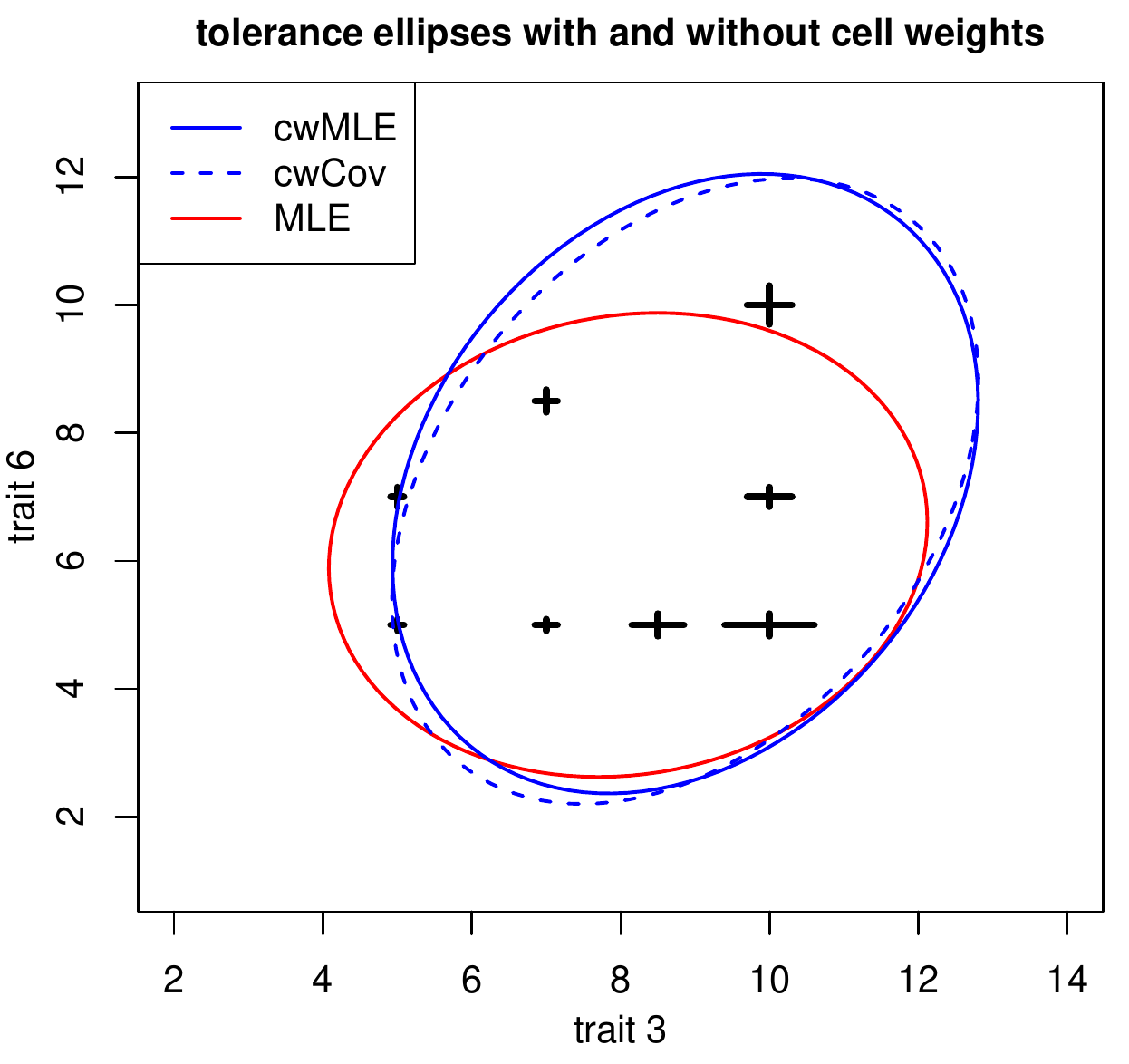}
\vspace{-0.2cm}	
\caption{Plot of variable t6 in 
  Table~\ref{exampledata} versus variable t3\,.
  The arms of the crosses reflect the cell 
	weight of each coordinate. 
	The 95\% tolerance ellipses
  of cwMLE, cwCov, and the unweighted MLE
  are shown.}
\label{fig:personality}				
\end{figure}

Figure~\ref{fig:personality} plots trait 6 
versus trait 3. 
The data points are shown as crosses, with 
the horizontal bar proportional to the cell
weight of the x-coordinate, and the vertical 
bar to that of the y-coordinate.
These weights vary a lot across the data.
We see only 8 crosses rather than 10 
because subjects 3 and 4 are tied here, as 
are subjects 5 and 9. We visualized this by
adding up the cell weights of the tied 
subjects.

The solid blue ellipse represents cwMLE,
and we see that its approximation cwCov 
(dashed line) is quite close to it.
Both are quite dissimilar to the red
ellipse of the unweighted MLE, which
extends further to the left and yields a
lower correlation coefficient (0.10
versus 0.32).
That the red ellipse extends further to
the left is because it gives all 
coordinates weight one, so the two 
x-coordinates on the left hand side pull 
as hard as all the others, unlike in 
cwMLE which takes their low cell weights
into account.
The centers of the blue ellipses lie
higher than the red one, and the blue 
ellipses are a bit slanted to the right, 
mainly due to the large vertical cell 
weight of the data point in the upper 
right corner.

\section{Summary and Outlook}

When faced with cellwise weighted data one
can use the proposed likelihood function, 
for which it is convenient to apply the 
unpacking transform to the data.
After this transform one can carry out
cellwise weighted maximum likelihood
estimation (cwMLE) of the parameters, as
well as likelihood-based inference.

For the ubiquitous multivariate gaussian 
model an iterative algorithm for the 
cwMLE is made available. 
The faster explicit methods
cwMean and cwCov are asymptotically
equivalent to the cwMLE and can be seen
as approximations, if needed after
regularizing cwCov to make it PSD.
In simulations the limiting behavior
was accurate already at relatively low
sample sizes.

A reviewer inquired about non-gaussian
data. The likelihood of an alternative
model distribution is different but 
formula~\eqref{eq:weightedfX} of 
the cellwise weighted likelihood 
can still be applied, as well as 
unpacking and the EM algorithm. 
For instance, the cwMLE can be used 
for data from a multivariate 
$t$-distribution, requiring only a bit 
more computation time.
The approximations cwMean and cwCov
are not as general and would obtain 
a lower statistical efficiency in 
that situation. Constructing fast 
approximations specifically tailored 
to the \mbox{$t$-distribution} would 
be harder since there is no explicit 
formula for its unweighted MLE to 
begin with.

The main benefit of this note is 
expected to be in the analysis of 
data in which the cells are measured
with different accuracies, or there 
are other reasons to assume that the
reliability varies across cells.
Section~\ref{sec:example} gave an
example with such data.
Other fields where data cells have
different accuracies are cDNA arrays 
\citep{lawrence2004} and
oligonucleotide arrays
\citep{turro2007}
where credibility intervals for the
data values are derived from 
posterior distributions.

Another type of application is to the
emerging field of cellwise outliers
that started with the publication of
\cite{alqallaf2009}.
There are methods that detect outlying
cells, such as the Detect Deviating 
Cells method \citep{DDC2018} or the 
cellwise MCD method \citep{cellMCD}. 
Both of these provide standardized
cellwise residuals, which are large
for outlying cells.
After such a method has run, one can
assign weights to the cells based on 
the size of their standardized 
cellwise residuals.
The approach proposed here can then
produce cellwise reweighted
estimates. This postprocessing step
may benefit the overall stability 
and accuracy of the final result.
It is analogous to the casewise 
reweighting step that is often carried 
out after a casewise robust method.\\

\noindent {\bf Software availability.}
An \texttt{R} implementation of the 
proposed techniques has been incorporated 
in the \texttt{cellWise} package 
\citep{cellWise} on CRAN, with the
vignette 
\texttt{cellwise\_weights\_examples}
reproducing the example in 
Section~\ref{sec:example}.\\

\newpage
\noindent {\bf Acknowledgment.}
Thanks go to Stefan Van Aelst, Jakob 
Raymaekers, and the reviewers for
helpful comments improving the
presentation.

\end{document}